\documentclass[preprint,1p,11pt]{IR-Template/ISAS_IR}

\usepackage[english]{babel}
\usepackage[utf8]{inputenc}
\usepackage{tikz}
\usepackage{caption}
\usepackage{subcaption}
\usepackage{booktabs}
\usepackage{setspace}
\usepackage{microtype}
\usepackage{enumerate}
\usepackage{url}

\usepackage{amsmath}
\usepackage{amssymb}
\usepackage{dsfont}
\usepackage{mathtools}
\usepackage{xspace}

\mathtoolsset{showonlyrefs}

\newcommand{\T}{^{\top}}

\newcommand{\dd}{\operatorname{d}\!}

\newcommand{\inv}[1]{\left(#1\right)^{-1}}
 
\DeclareMathOperator{\diag}{diag}

\newcommand{\idx}[1]{^{(#1)}}
\newcommand{\idxi}{\idx{i}}
\newcommand{\idxj}{\idx{j}}

\newcommand{\mean}[1]{\hat{#1}}

\newcommand{\Gauss}{\mathcal{N}}

\renewcommand{\vec}[1]{\underline{#1}}

\newcommand{\va}{\vec{a}}
\newcommand{\vak}{\va_{k}}

\newcommand{\vg}{\vec{g}}

\newcommand{\vh}{\vec{h}}

\newcommand{\vm}{\vec{m}}
\newcommand{\vmk}{\vm_{k}}

\newcommand{\vo}{\vec{o}}
\newcommand{\vok}{\vo_{k}}
\newcommand{\evo}{\mean{\vo}}

\newcommand{\vr}{\vec{r}}
\newcommand{\vrk}{\vr_{k}} 
\newcommand{\evr}{\mean{\vr}}

\newcommand{\vv}{\vec{v}}
\newcommand{\vvk}{\vv_{k}}

\newcommand{\vw}{\vec{w}}
\newcommand{\vwk}{\vw_{k}}

\newcommand{\vx}{\vec{x}}
\newcommand{\vxk}{\vx_{k}}
\newcommand{\vxkk}{\vx_{k-1}}

\newcommand{\evx}{\mean{\vx}}

\newcommand{\evxkp}{\evx_{k}^{p}}
\newcommand{\evxke}{\evx_{k}^{e}}
\newcommand{\evxkke}{\evx_{k-1}^{e}}

\newcommand{\vtheta}{\vec{\theta}}
\newcommand{\vthetak}{\vtheta_k}

\newcommand{\vTheta}{\vec{\Theta}}
\newcommand{\vThetak}{\vTheta_k}
\newcommand{\evTheta}{\mean{\vTheta}}

\newcommand{\mat}[1]{\mathbf #1}

\newcommand{\mC}{\mat{C}}

\newcommand{\mI}{\mat{I}}

\newcommand{\mQ}{\mat{Q}}
\newcommand{\mQk}{\mQ_{k}}

\newcommand{\mR}{\mat{R}}
\newcommand{\mRk}{\mR_{k}}
\newcommand{\mRki}{\mR_{k}\idxi}

\newcommand{\Cr}{\mC^{r}}

\newcommand{\Co}{\mC^{o}}

\newcommand{\Cxkp}{\mC_{k}^{p}}
\newcommand{\Cxke}{\mC_{k}^{e}}
\newcommand{\Cxkke}{\mC_{k-1}^{e}}

\newcommand{\CTheta}{\mC^{\Theta}}

\newcommand{\vmeas}{\tilde{\vec{m}}}
\newcommand{\vmeask}{\vmeas_{k}}

\newcommand{\measset}{\mathcal{M}}
\newcommand{\meassetk}{\measset_{k}}

\newcommand{\sskf}{S\textsuperscript{2}KF\xspace}

\newcommand{\Bmat}{\begin{bmatrix}}
\newcommand{\Emat}{\end{bmatrix}}

\newcommand{\Beq}{\begin{equation}\begin{aligned}}
\newcommand{\Eeq}{\end{aligned}\end{equation}}

\newcommand{\Sec}[1]{Sec.~\ref{#1}}             
\newcommand{\Eq}[1]{\eqref{#1}}                 
\newcommand{\Fig}[1]{Fig.~\ref{#1}}             

\begin{document}
    \begin{frontmatter}
        \title{High-Accuracy Real-Time\\Whole-Body Human Motion Tracking\\Based on Constrained Nonlinear Kalman Filtering}
        
        \author[isas]{Jannik~Steinbring}
        \ead{jannik.steinbring@kit.edu}
        
        \author[h2t]{Christian~Mandery}
        \ead{mandery@kit.edu}
        
        \author[h2t]{Nikolaus~Vahrenkamp}
        \ead{vahrenkamp@kit.edu}
        
        \author[h2t]{Tamim~Asfour}
        \ead{asfour@kit.edu}
        
        \author[isas]{Uwe~D.~Hanebeck}
        \ead{uwe.hanebeck@ieee.org}
        
        \address[isas]{Intelligent Sensor-Actuator-Systems Laboratory (ISAS)\\
                       Institute for Anthropomatics and Robotics\\
                       Karlsruhe Institute of Technology (KIT), Germany\vspace{3mm}}
        
        \address[h2t]{High Performance Humanoid Technologies Lab (H\textsuperscript{2}T)\\
                      Institute for Anthropomatics and Robotics\\
                      Karlsruhe Institute of Technology (KIT), Germany\vspace{3mm}}

\begin{abstract}

We present a new online approach to track human whole-body motion from motion capture data, i.e., positions of labeled markers attached to the human body.
Tracking in noisy data can be effectively performed with the aid of well-established recursive state estimation techniques.
This allows us to systematically take noise of the marker measurements into account.
However, as joint limits imposed by the human body have to be satisfied during estimation, first we transform this constrained estimation problem into an unconstrained one by using periodic functions.
Then, we apply the Smart Sampling Kalman Filter to solve this unconstrained estimation problem.
The proposed recursive state estimation approach makes the human motion tracking very robust to partial occlusion of markers and avoids any special treatment or reconstruction of the missed markers.
A concrete implementation built on the kinematic human reference model of the Master Motor Map framework and a Vicon motion capture system is evaluated.
Different captured motions show that our implementation can accurately estimate whole-body human motion in real-time and outperforms existing gradient-based approaches.
In addition, we demonstrate its ability to smoothly handle incomplete marker data.

\end{abstract}

    \end{frontmatter}

\section{Introduction}
\label{sec:introduction}

Understanding human whole-body motion has been a fundamental research interest with numerous applications in the robotics community.
Great efforts have been done to establish procedures for capturing, representation, processing, and transfer of human motion in robotics, such as the Master Motor Map (MMM) framework \cite{Terlemez2014_MMM} providing a unifying reference model of the human body, which this work builds upon.

\begin{figure}
    \centering
    \includegraphics[width=0.25\textwidth]{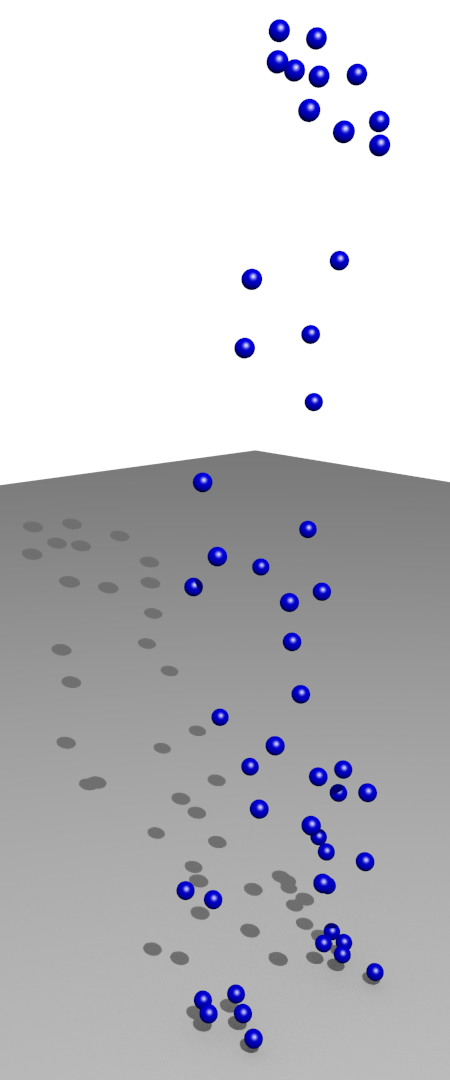}
    \includegraphics[width=0.25\textwidth]{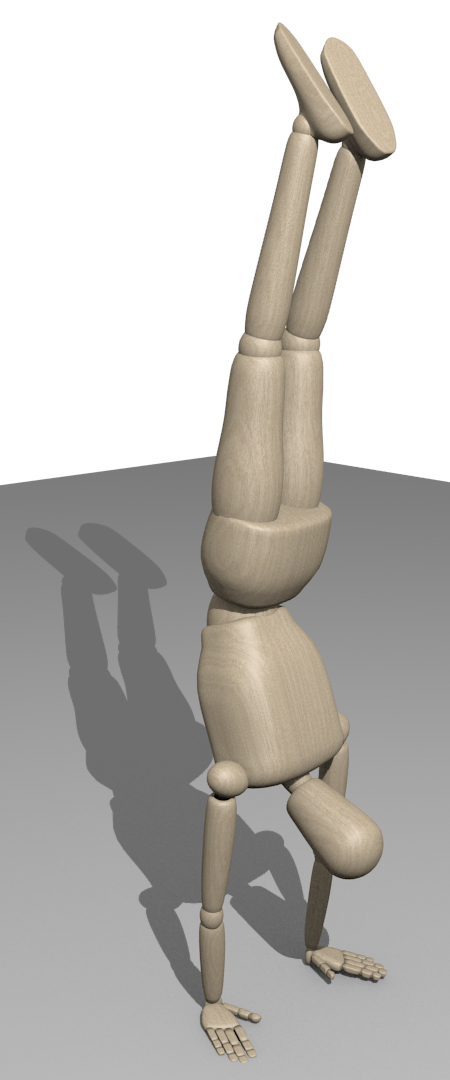}
    \caption{Single frame of a tracked motion from labeled markers. Measured markers on the left. Pose estimated with the proposed approach on the right.}
    \label{fig:face}
\end{figure}

Today, several commercial systems offer an easy way to capture human motion and provide accurate Cartesian measurements of labeled markers attached to the human body.
Based on those measurements, the whole-body human motion can be tracked by estimating certain parameters of a given kinematic model, e.g., root pose and joint angles (see \Fig{fig:face}).
However, like all measurements, the captured marker positions suffer from noise.
Hence, stochastic approaches are needed to obtain a precise estimation of all the kinematic parameters of interest.
Additionally, joint limits imposed by the human body have to be satisfied during estimation.

Tracking the human motion is equivalent to estimating the state of a discrete-time stochastic nonlinear dynamic system, where the system state is the human pose.
This allows us to employ a recursive nonlinear state estimator to solve the human motion tracking problem.
The advantage of such an estimator is that it maintains a probability distribution of  the state estimate rather than only a simple set of values.
The estimator uses this distribution to optimally fuse the current state estimate with newly available noisy measurements to obtain an updated distribution.
Popular state estimators are (nonlinear) Kalman filters \cite{kazufumi_ito_gaussian_2000, simon_j._julier_unscented_2004, dan_simon_optimal_2006, pawe_stano_parametric_2013} or particle filters \cite{branko_ristic_beyond_2004, arnaud_doucet_tutorial_2011}.
Due to the many degrees of freedom (DoF) required for a detailed human pose, particle filters are not suitable for real-time tracking  as they would need a huge amount of particles to get meaningful state estimates.
Hence, we choose to use the Smart Sampling Kalman Filter (\sskf) \cite{jannik_steinbring_lrkf_2014} to estimate all parameters of the kinematic model in real-time.
Furthermore, we transform the constrained estimation problem into an unconstrained problem with the aid of periodic functions.
This is necessary to satisfy all the imposed joint limits as Kalman filters can only estimate unconstrained quantities.
Furthermore, the problem of missing marker positions, e.g., due to occlusions, is addressed.
Thanks to the recursive state estimation approach, it is possible to ignore missing marker measurements while still obtain good tracking results based on the remaining measurements only.

In \cite{johannes_meyer_online_2014}, the authors solve the more general human motion tracking problem consisting of unlabeled markers and an a priori unknown skeleton that has to be fitted to that marker data.
Here, however, we assume that the marker associations are already known and that we can make use of knowledge about the kinematic model to get highly accurate estimates.
Therefore, their special initialization phase, i.e., a T-pose performed by the human, is not required by our approach.
Moreover, the authors are not clear about the real-time capabilities of their approach.

The problem of missing marker positions is addressed by the authors of \cite{andreas_aristidou_real-time_2013}.
Their solution is to predict missing marker positions, use previously known marker positions, and get information based on rigid body assumptions.
Nonetheless, our proposed recursive state estimation approach implicitly takes information of previous frames into account to be able to handle missing markers more easily.
Moreover, their tracking approach does not work with a fixed kinematic model.
That is, they do online joint localization in the marker point cloud, which results in a time-varying kinematic model.

A force-based approach in the fields of computer graphics is taken in \cite{victor_b._zordan_mapping_2003}.
The authors solve the tracking problem with a physical simulation.
Unfortunately, their approach does not consider the noise of the measurements.

The remainder of this paper is structured as follows.
In the next Section, we briefly present the aforementioned MMM framework.
After that, in \Sec{sec:estimation}, we present a general way to track whole-body human motion over time from noisy marker measurements using a constrained nonlinear Kalman filter.
Based on the MMM framework, we formulate a concrete implementation of this general tracking approach in \Sec{sec:converter}.
We evaluate the implementation with a complex motion in \Sec{sec:evaluation}.
Finally, the conclusions are given in \Sec{sec:conclusions}.

\section{The Master Motor Map Framework}
\label{sec:mmm}

\begin{figure}
    \centering
    \includegraphics[width=0.35\textwidth]{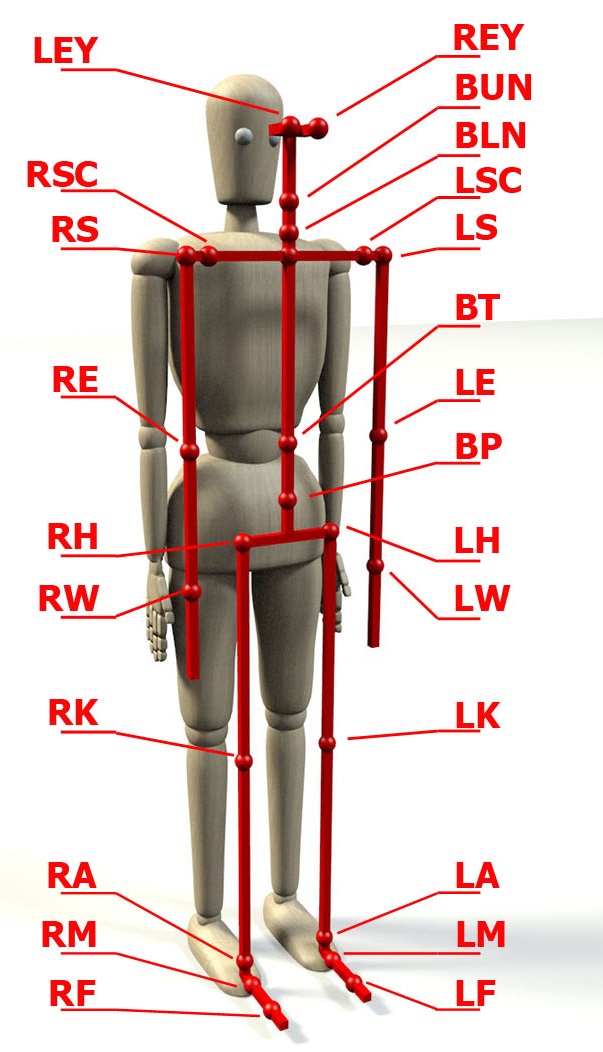}
    \caption{Whole-body kinematics of the MMM reference model, including joints abbreviations (from~\cite{Terlemez2014_MMM}).}
    \label{fig:mmm-kinematics}
\end{figure}

The Master Motor Map (MMM) framework \cite{Azad2007_MMM, Terlemez2014_MMM} provides an open-source framework for the capturing, representation and analysis of human motion, and its reproduction on humanoid robots.
It has been used in a number of different robotics research applications that leverage human motion, e.g. \cite{Do2011_GraspRepresentation, Asfour2011_SensorimotorPrimitives, Mandery2015_WholebodyPoseTransitions}.
At its core, the MMM framework provides the MMM reference model, a whole-body model containing both kinematic and dynamic specifications for the human body based on well-established biomechanical literature by Winter \cite{Winter2009_Biomechanics} and others.
This reference model allows the representation of human motion using 6 DoF for the root pose, 52 DoF for torso, extremities, head, and eyes, and 2 $\times$ 23 DoF for both hands.
\Fig{fig:mmm-kinematics} shows the kinematics of the MMM reference model.

A central proposition of the MMM approach is to use the MMM reference model as a unique intermediate model that allows the unifying representation of human motion provided by different motion input sources, e.g., marker-less or marker-based motion capture, visual approaches based on 2D or depth images, or data captured from inertial measurement units.
Therefore, procedures are needed that allow to transfer raw input data to the kinematic embodiment of the MMM model, reconstructing position, orientation, and joint angle values of the model.
In the terminology of the MMM, these modules are called \emph{converters}.

For the use of motion capture data as the input source, the MMM framework provides procedures for the recording of motion that include a marker set for human whole-body motion capture comprising a total of 56 marker locations at characteristic anatomical landmarks of the human body\footnote{In-depth specifications of the marker set are available online: \url{https://motion-database.humanoids.kit.edu/marker_set/}}.
Virtual markers that match the markers placed on the subject are then added to the MMM reference model at the corresponding locations.
One converter is already provided by the MMM framework for the reconstruction of human motion from motion capture \cite{Terlemez2014_MMM}.
This converter uses a framewise gradient-based optimization approach based on the Jacobian matrix of the MMM kinematics to fit the model pose and is compared to the new approach proposed in this paper in \Sec{sec:evaluation}.

\section{Whole-Body Human Motion Tracking}
\label{sec:estimation}

In this Section, we present a new way to track the motion of a human based on noisy marker measurements using a constrained nonlinear Kalman filter.
The proposed approach is not limited to a certain kinematic model of the human motion or how the measurements of the markers are obtained.

\subsection{Problem Formulation}

Given a kinematic model of the human body parameterized by $J$ joint angles
\Beq
    \vthetak &= [\theta_k\idx{1}, \ldots, \theta_k\idx{J}]\T
\Eeq
and root pose consisting of position
\Beq
    \vrk &= [r_k^x, r_k^y, r_k^z]\T
\Eeq
in Cartesian space and orientation
\Beq
    \vok &= [o_k^r, o_k^p, o_k^y]\T
\Eeq
in roll, pitch, and yaw angles\footnote{Vectors are underlined and matrices are printed bold face.}, our goal is to estimate the pose of a human at discrete time steps $k$.
The estimation relies on several unique markers attached to the human body at \emph{known} positions, e.g., on a shoulder or hand that are observed and measured by a tracking system.
At each time step $k$, the tracking system provides us with a set
\Beq
    \meassetk = \{ \vmeask\idx{1}, \ldots, \vmeask\idx{M} \}
\Eeq
of $M$ labeled noisy marker positions $\vmeask\idxi$ in Cartesian space.
In addition, due to human joint limitations, the estimated pose has to satisfy the bound constraint
\Beq
    \label{eq:joint-constraint}
    l_j \leq \theta_k\idxj \leq u_j \quad \forall j \in \{ 1, \ldots, J \}
\Eeq
for all joint angles~$\theta_k\idxj$ with lower bound~$l_j$ and upper bound~$u_j$.

\subsection{From Model Parameters to Marker Positions}

In order to infer the kinematic model parameters $\vthetak$, $\vrk$, and $\vok$ from the received marker positions $\meassetk$ based on a nonlinear Kalman filter, we need a mapping from these parameters to each individual marker position.
This mapping consists of two parts.
First, given a concrete pose (in form of the parameters $\vthetak$, $\vrk$, and $\vok$), for each marker the forward kinematics of the respective kinematic chain has to be computed.
As a result, it is known where to \emph{expect} all markers if the human has that concrete pose.
Second, like all measurements, the measured marker positions suffer from noise.
Hence, that noise has to be taken into account in order to obtain good estimation results, especially in case of strong noise.
Both together leads to the desired mapping
\Beq
    \label{eq:meas-model}
    \vmk\idxi = \vh\idxi(\vrk, \vok, \vthetak) + \vvk\idxi \enspace,
\Eeq
where $\vmk\idxi$ denotes the $i$-th marker position in Cartesian space, $\vh\idxi(\cdot, \cdot, \cdot)$ the forward kinematics for the $i$-th marker, and $\vvk\idxi$ additive, zero-mean, and white Gaussian noise with covariance matrix $\mRki$.
The choice of $\mRki$ depends on the utilized tracking system.
Moreover, it is assumed that the noise vectors $\vvk\idxi$ and $\vvk\idxj$ with $i \neq j$ are mutually independent.
Note also the difference between $\vmk\idxi$ and $\vmeask\idxi$: the first one is a random vector, whereas the latter one is a \emph{realization} of that random vector.

\subsection{Satisfying the Joint Angle Bound Constraints}

The considered estimation task poses an additional challenge for Kalman filters: a (nonlinear) Kalman filter, by design, only estimates unconstrained quantities.
That is, directly estimating $\vthetak$ with a Kalman filter will violate the constraints~\Eq{eq:joint-constraint}.
Recall that the estimate of a Kalman filter is represented by a mean vector and a covariance matrix.
In order to take the bound constraints properly into account, it is necessary that (i) the mean vector must always lie inside the bounded region of the state space, and (ii) the covariance matrix also has to reflect that the state space is bounded, that is, the covariance matrix has to be smaller compared to an unconstrained state space.
In literature, there exist various approaches to incorporate constraints into Kalman filters.
\begin{itemize}
  \item
\emph{Perfect measurements} \cite{dan_simon_optimal_2006} are designed for equality constraints and are not suitable for inequality constraints.
Hence, they cannot be applied to the considered bound constraint problem.
  \item
\emph{Projection techniques} \cite{dan_simon_optimal_2006} correct the posterior state mean after a Kalman filter prediction/update step.
Unfortunately, they cannot correct the posterior state covariance matrix as well.
  \item
\emph{Pdf truncation} \cite{dan_simon_optimal_2006} is an elegant way to respect linear inequality constraints and corrects both posterior state mean and covariance matrix.
However, it is computationally expensive for large state dimensions as it requires several Gram-Schmidt orthogonalizations and eigendecompositions of the state covariance matrix, which is not guaranteed to converge, and hence, makes this approach unreliable.
  \item
The \emph{sampling-based approach} proposed in \cite{ondrej_straka_truncation_2012} can be seen as a numerical approximation of the pdf truncation approach.
The problem here is that situations may occur where all samples lie outside of the constrained region and no constrained estimate can be obtained.
This is analogous to the known sample degeneracy problem of particle filters.
\end{itemize}
As we seek a real-time capable and accurate human motion tracking, we choose another way to satisfy \Eq{eq:joint-constraint} for all joint angles.
We perform a parameter transformation using a periodic function that is defined on $(-\infty, \infty)$ but its range is limited to the interval $[-1, 1]$.
We introduce a new joint parameter $\Theta_k\idx{j}$ for each joint angle $\theta_k\idx{j}$ according to the mapping
\Beq
    \theta_k\idxj = g_j(\Theta_k\idxj) = \frac{u_j - l_j}{2} \sin(\Theta_k\idxj) + \frac{l_j + u_j}{2} \enspace.
\Eeq
As a result, $\Theta_k\idxj$ can take any value, i.e., it is unconstrained, while \Eq{eq:joint-constraint} is always satisfied.
It should be noted that this periodic approach, however, is sensitive to large uncertainties in the parameters $\Theta_k\idxj$, that is, its uncertainty should not be larger than the period of the periodic function to get meaningful estimation results.

Alternatively, sigmoid functions like the hyperbolic tangent could also be used for such a transformation.
However, a nonlinear Kalman filter has problems to properly update a joint angle estimate in situations where it is near to a bound constraint as the gradient of a sigmoid function becomes very small for large parameters.

Analogously to the vector $\vthetak$, we define the joint parameter vector
\Beq
    \vThetak = [\Theta_k\idx{1}, \ldots, \Theta_k\idx{J}]\T \enspace.
\Eeq
We also introduce the vector-valued function
\Beq
    \label{eq:transformation-func}
    \vthetak = \vg(\vThetak) = \Bmat g_{1}(\Theta_k\idx{1}) \\
                                     \vdots \\
                                     g_{J}(\Theta_k\idx{J}) \Emat
\Eeq
that transforms all joint parameters back to their corresponding joint angles.

\subsection{State Estimation with the Smart Sampling Kalman Filter}

At this point, we can introduce the system state
\Beq
    \label{eq:sys-state}
    \vxk = [\vrk\T, \vok\T, \vThetak\T]\T
\Eeq
that fully describes the constrained whole-body human motion at time step~$k$.
This state vector can now be estimated with a usual nonlinear Kalman filter.
A Kalman filter is a recursive state estimator consisting of two alternating parts: (i) the prediction step that propagates the state estimate, that is, mean and covariance matrix, from the last time step~$k - 1$ to the current time step~$k$ and (ii) the measurement update that corrects the predicted state estimate given a set of measurements.

First, we concentrate on the measurement update.
Here, we have to define what the actual measurement is that will be processed by the Kalman filter, and the measurement equation that maps $\vxk$ to that measurement.
To obtain the measurement, we stack the received single marker measurements $\vmeask\idx{i}$ to a $3M$-dimensional measurement vector
\Beq
    \label{eq:stacked-meas}
    \vmeask = [(\vmeask\idx{1})\T, \ldots, (\vmeask\idx{M})\T]\T \enspace.
\Eeq
For the measurement equation, we combine the $M$ marker mappings~\Eq{eq:meas-model} with \Eq{eq:transformation-func},
and stack them in the same manner of \Eq{eq:stacked-meas} which yields
\Beq
    \label{eq:stacked-meas-model}
    \underbrace{\Bmat \vmk\idx{1} \\ \vdots \\ \vmk\idx{M} \Emat}_{\vmk} =
    \underbrace{\Bmat \vh\idx{1}(\vrk, \vok, \vg(\vThetak)) \\[0.12cm] \vdots \\[0.12cm] \vh\idx{M}(\vrk, \vok, \vg(\vThetak)) \Emat}_{\vh(\vxk)} +
    \underbrace{\Bmat \vvk\idx{1} \\[0cm] \vdots \\[0cm] \vvk\idx{M} \Emat}_{\vvk} \enspace,
\Eeq
where the zero-mean noise vector $\vvk$ has the block diagonal covariance matrix
\Beq
    \mRk = \diag(\mRk\idx{1}, \ldots, \mRk\idx{M}) \enspace.
\Eeq
Of course, the order of stacking has to be the same for both \Eq{eq:stacked-meas} and \Eq{eq:stacked-meas-model}.
Otherwise, marker positions would not be related to their corresponding measurements.
Now, given a predicted state estimate by state mean $\evxkp$ and state covariance matrix $\Cxkp$, the following moments are computed based on the measurement model \Eq{eq:stacked-meas-model}
\Beq
    \label{eq:meas-moments}
    \mean{\vm}_k &= \int \vh(\vxk) \cdot \Gauss(\vxk; \evxkp, \Cxkp) \dd \vxk \\
    \mC^m_k      &= \int (\vh(\vxk) - \mean{\vm}_k) \cdot (\vh(\vxk) - \mean{\vm}_k)\T \cdot \\
                 &\hspace*{0.9cm} \Gauss(\vxk; \evxkp, \Cxkp) \dd \vxk + \mRk \\
    \mC^{x,m}_k  &= \int (\vxk - \evxkp) \cdot (\vh(\vxk) - \mean{\vm}_k)\T \cdot \\
                 &\hspace*{0.9cm} \Gauss(\vxk; \evxkp, \Cxkp) \dd \vxk \enspace,
\Eeq
where $\Gauss(\vxk; \evxkp, \Cxkp)$ denotes the Gaussian probability density function.

Together with the measurement~\Eq{eq:stacked-meas}, we get the posterior, i.e., the corrected, state estimate by computing the posterior state mean according to
\Beq
    \label{eq:updated-state-mean}
    \evxke &= \evxkp + \mC^{x,m}_k \inv{\mC^m_k} (\vmeask - \mean{\vm}_k) \enspace,
\Eeq
and the posterior state covariance matrix according to
\Beq
    \label{eq:updated-state-cov}
    \Cxke &= \Cxkp - \mC^{x,m}_k \inv{\mC^m_k} (\mC^{x,m}_k)\T \enspace.
\Eeq

Second, the prediction step requires a system equation that models the temporal evolution of the system state between two measurement updates.
Here, this means (slight) changes in the human motion from one time step to the next one, e.g., a movement of the root position or moving an extremity.
A general system equation is given by
\Beq
    \label{eq:sys-model}
    \vxk = \vak(\vxkk) + \vwk \enspace,
\Eeq
where $\vwk$ denotes zero-mean white Gaussian noise with covariance matrix $\mQk$.
The function $\vak(\cdot)$ models the actual changes in the kinematic model parameters over time and should take any prior knowledge about the human motion scenario into account.
For example, it can rely on velocities to predict where the human or its extremities will be in the next time step.
Note that such velocities can be estimated together with the actual kinematic model parameters by augmenting the state system~\Eq{eq:sys-state} with such velocities.
The noise, and therefore $\mQk$, incorporates modelling errors into the prediction and depends on the used tracking system and the elapsed time between measurement updates.
Given the system equation \Eq{eq:sys-model} and the state estimate from the last time step by state mean $\evxkke$ and state covariance matrix $\Cxkke$, the Kalman filter computes the predicted state mean according to
\Beq
    \label{eq:pred-state-mean}
    \evxkp &= \int \vak(\vxkk) \cdot \Gauss(\vxkk; \evxkke, \Cxkke) \dd \vxkk \enspace,
\Eeq
and the predicted state covariance matrix according to
\Beq
    \label{eq:pred-state-cov}
    \Cxkp &= \int (\vak(\vxkk) - \evxkp) \cdot (\vak(\vxkk) - \evxkp)\T \cdot \\
          &\hspace*{0.9cm} \Gauss(\vxkk; \evxkke, \Cxkke) \dd \vxkk + \mQk \enspace.
\Eeq

Up to this point, no concrete nonlinear Kalman filter has been chosen to compute the multi-dimensional integrals in \Eq{eq:meas-moments}, \Eq{eq:pred-state-mean}, and \Eq{eq:pred-state-cov}.
Basically, every nonlinear Kalman filter such as the extended Kalman filter (EKF) \cite{dan_simon_optimal_2006} or the unscented Kalman filter (UKF) \cite{simon_j._julier_unscented_2004} could be used.
However, on the one hand, the EKF relies on explicit linearization of the measurement model around the predicted state mean, which requires the Jacobian matrix of \Eq{eq:stacked-meas-model} and \Eq{eq:sys-model}.
Moreover, it does not incorporate the uncertainty of the predicted state estimate into this linearization, making this approach sensitive to the predicted state mean.
On the other hand, the UKF relies on statistical linearization, which incorporates the prior uncertainty and does not need any Jacobian matrix.
Instead, the UKF propagates a set of samples through the measurement/system model to compute the integrals.
Unfortunately, the number of samples is fixed and cannot be increased to obtain more reliable and more accurate state estimates.
To get rid of the limitations imposed by the EKF and UKF, we use the Smart Sampling Kalman Filter (\sskf) \cite{jannik_steinbring_s2kf:_2013, jannik_steinbring_lrkf_2014}.
The \sskf also relies on samples to compute the integrals, but it can use an arbitrary number of optimally placed samples\footnote{An open-source MATLAB implementation of the \sskf is available online: \url{https://bitbucket.org/nonlinearestimation/toolbox/}}.
Hence, we can use more samples than the UKF to improve the estimation quality, but only as many as possible to guarantee real-time capability.  

Finally, to start with the recursive state estimation, an initial state estimate with initial mean $\evx_0^e$ and initial covariance matrix $\mC_0^e$ is required.
These initial values depend on the quality of the measurements and other prior knowledge of the human motion scenario, e.g., if it is known where the human starts or what its initial pose is.

\subsection{Working with Incomplete Measurement Sets}
\label{sec:incomplete-meas-sets}

If the position $\vmeask\idx{i}$ for the $i$-th marker cannot be obtained by the tracking system at time step $k$, e.g., due to occlusions, we omit it from \Eq{eq:stacked-meas} and the corresponding measurement function $\vh\idx{i}(\cdot)$ from \Eq{eq:stacked-meas-model}, and use only the remaining measured marker positions for the measurement update.
That is, an estimation of root pose and joint angles is still possible for that time step.
However, due to the lack of certain marker position measurements, the filter has less information about the current human pose.
Consequently, the estimation quality can be less accurate.
Nonetheless, an estimation is still possible thanks to prior knowledge of the pose, i.e., the predicted state estimate.

\section{A New Converter for the MMM Framework}
\label{sec:converter}

After describing the general whole-body human motion tracking approach based on constrained nonlinear Kalman filtering in \Sec{sec:estimation}, we implement this approach for the kinematic model presented in \Sec{sec:mmm}.
Here, we select a subset of the joints from the MMM reference model based on the parts of motion that can be estimated using our motion capture setup.
That is, hands, eyes, and some joints on shoulders and feet have been excluded.
In total, $J = 40$ joints are used for the kinematic model resulting in a system state dimension of $46$.
Moreover, root position and marker positions are measured in millimeters and root orientation and joint angles are measured in radians (this is important as it also defines the units of the noise covariance matrices $\mRki$ and $\mQk$).

In addition, the MMM marker set describes the placement on markers on the human body and the MMM reference model provides the corresponding forward kinematics $\vh\idxi(\cdot, \cdot, \cdot)$ required for the measurement model \Eq{eq:stacked-meas-model}.
We use $M = 50$ of those markers.
Their positions are measured with a Vicon MX10 system using ten T10 cameras.
It is an optical motion capture system based on passive (reflective) markers.
The system records at 100~Hz, that is, every 10~ms we get a new set of markers $\meassetk$.
For the measurement update, the measurement noise properties of the Vicon system, i.e., the covariance matrices $\mRki$, have to be known.
Experimentally, we have found that the marker positions provided by the Vicon system are disturbed approximately with
\Beq
    \mRki = 10^{-4}\mI_3 \enspace,
\Eeq
where $\mI_3$ denotes the identity matrix of dimension three.
To perform the measurement update, the \sskf is configured to use 301 samples.
This is more than six times the number of samples that would be used by the UKF.

The highly accurate marker position measurements provided by the Vicon system allows us to model the temporal evolution of the human motion with a simple random walk system model according to
\Beq
    \vxk = \vxkk + \vwk \enspace.
\Eeq
That is, it is assumed that the system state does not evolve significantly in the 10~ms between two measurement updates (only the uncertainty of the estimate will increase) and that the measurement update corrects the state estimate adequately.
As a consequence, the Kalman filter prediction can be computed easily in closed-form with
\Beq
    \evxkp &= \evxkke \\
    \Cxkp  &= \Cxkke + \mQk \enspace,
\Eeq
that is, no sampling (like for the measurement update) is required at all.
Furthermore, the system noise covariance matrix is set to the time-invariant diagonal matrix
\Beq
    \mQk = \diag(25, 25, 25, 10^{-10}, \ldots, 10^{-10}) \enspace.
\Eeq

For the initial system state estimate, we assume no prior knowledge about the human motion scenario.
This makes this implementation very versatile to track any human motion without a special treatment.
More precisely, we use the first set of available measurements $\measset_0$ to initialize the \sskf.
The root position and its covariance is obtained according to
\Beq
     \evr_0 &= \frac{1}{M} \sum_{i=1}^M \vmeas_0\idxi \enspace,\\
     \Cr_0  &= \frac{1}{M} \sum_{i=1}^M (\vmeas_0\idxi - \evr_0) \cdot (\vmeas_0\idxi - \evr_0)\T \enspace,
\Eeq
the orientation mean and covariance is set to
\Beq
    \evo_0 &= [0, 0, 0]\T \enspace,\\
    \Co_0  &= 10^{-6} \mI_3 \enspace,
\Eeq
and the joint parameters and their uncertainties are initialized with
\Beq
    \evTheta_0 &= [0, \ldots, 0]\T \enspace,\\
    \CTheta_0  &= 10^{-10} \mI_{40} \enspace,
\Eeq
that is, each initial joint angle is set to the average of its bound constraints.
Then, the initial system state estimate is given by
\Beq
    \evx_0^e &= [\evr_0\T, \evo_0\T, \evTheta_0\T]\T \enspace,\\
    \mC_0^e  &= \diag(\Cr_0, \Co_0, \CTheta_0) \enspace.
\Eeq

\section{Evaluation}
\label{sec:evaluation}

In this Section, we evaluate the implemented human motion tracking approach from \Sec{sec:converter} with a performed handstand.
Its Vicon motion recording was taken from the KIT Whole-Body Human Motion Database \cite{Mandery2015_MotionDB}, which provides a rich collection of raw human motion capture recordings for numerous kinds of motion tasks in the industry standard C3D file format.
The recording contains 675 frames (time steps).
In all frames, the entire set of marker positions is available.
That means that we can work with a complete measurement set in all time steps.

We compare the new approach with the Jacobian-based MMM converter \cite{Terlemez2014_MMM} mentioned in \Sec{sec:mmm}.
In order to assess the human motion tracking performance, for each estimated pose (one for each time step), we compute the expected marker positions using their respective forward kinematics.
Then, we compute the distances between the expected and the measured marker positions and subsequently build the average over all these distances.

\begin{figure}
    \centering
    \includegraphics[width=0.55\textwidth]{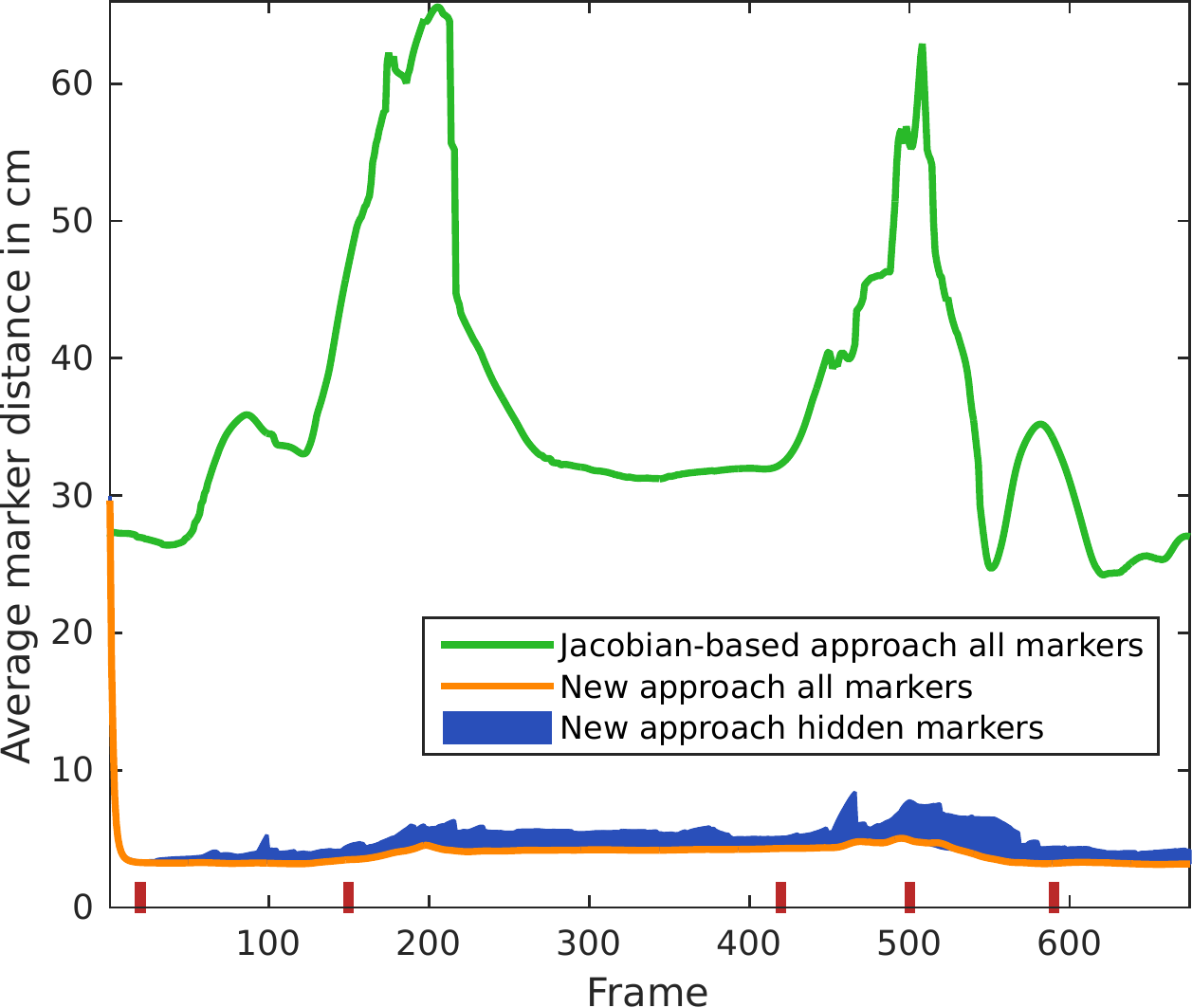}
    \caption{Averaged distances between measured marker positions and marker positions expected by the estimated pose. The red lines indicate the frames shown in Figures \ref{fig:handstand-no-gaps} and \ref{fig:handstand-gaps}.}
    \label{fig:handstand-marker-distances}
\end{figure}

The marker distances for the new approach (orange) and the Jacobian-based approach (green) are depicted in \Fig{fig:handstand-marker-distances}.
On the one hand, it can be seen that the new approach requires approximately 20 frames (only 200~ms) to converge.
This can be explained with the nonlinear Kalman filter that gradually improves its initial state estimate over time.
However, after convergence, the marker distances of the new approach do not change much over time, even when the pose changes drastically at beginning and end of the handstand.
It offers a total average marker distance of only 4 to 5~cm.
It is important to note that a further reduction in the marker distances is not straightforward.
The problem is that markers attached to the human body will never exactly coincide with the corresponding marker positions defined in the reference model.
On the other hand, the Jacobian-based approach has problems to track the handstand motion.
Although it can handle ordinary motions, like human bipedal locomotion tasks well, it clearly has problems with such a complex motion as the handstand.
At beginning and end of the handstand, its average marker distances are over 60~cm.
Also the general marker distance level is much higher compared to the new approach.

\begin{figure}
    \centering
    \includegraphics[width=0.55\textwidth]{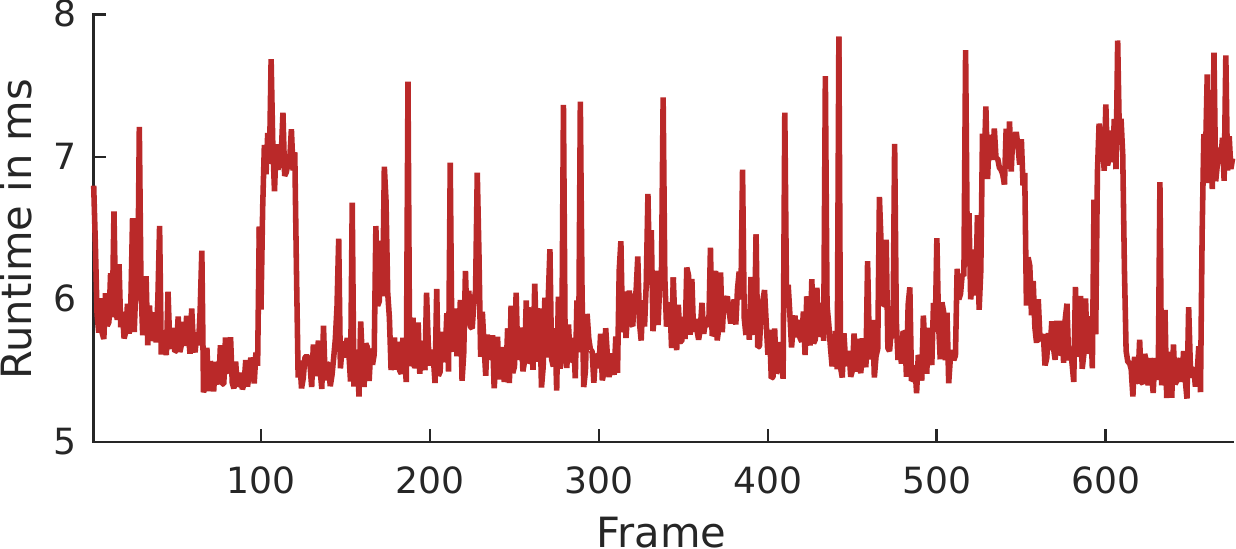}
    \caption{Estimation runtime of the new approach. The evaluation is performed on an Intel Core i7-3770 CPU.}
    \label{fig:handstand-no-gaps-runtimes}
\end{figure}

\begin{figure*}
    \centering
    \includegraphics[width=0.19\textwidth]{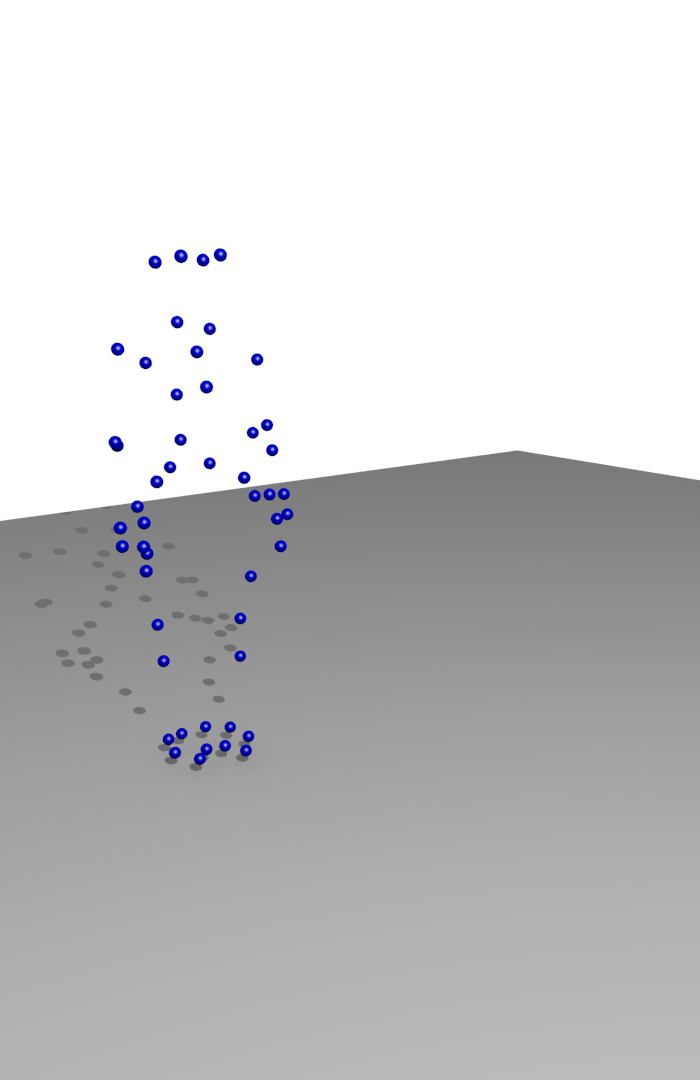}
    \includegraphics[width=0.19\textwidth]{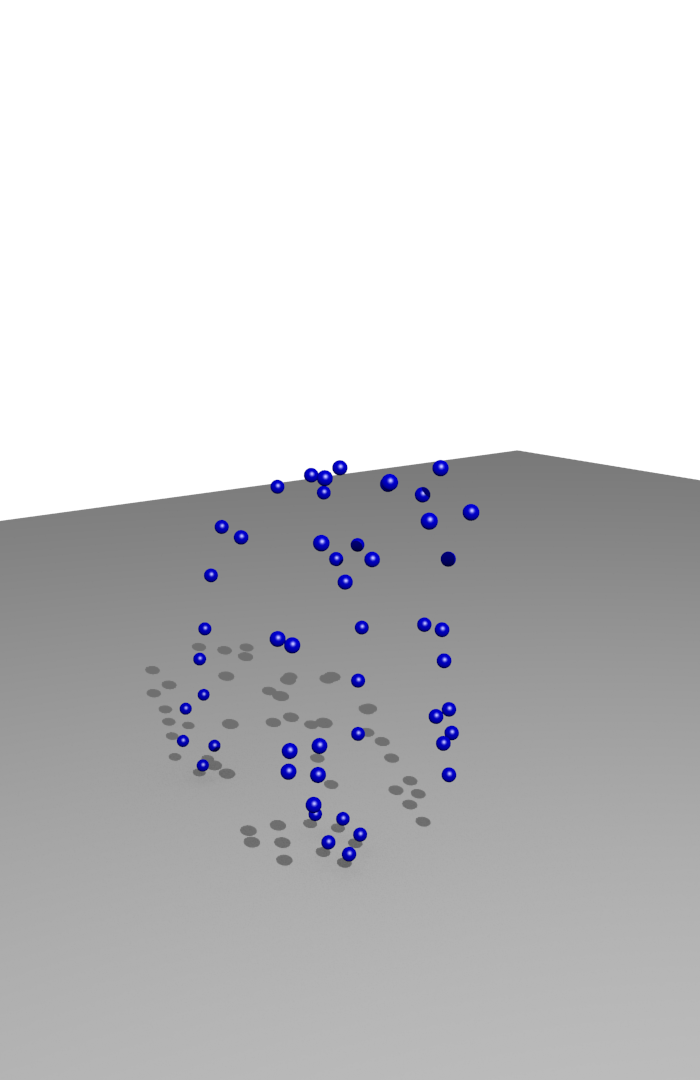}
    \includegraphics[width=0.19\textwidth]{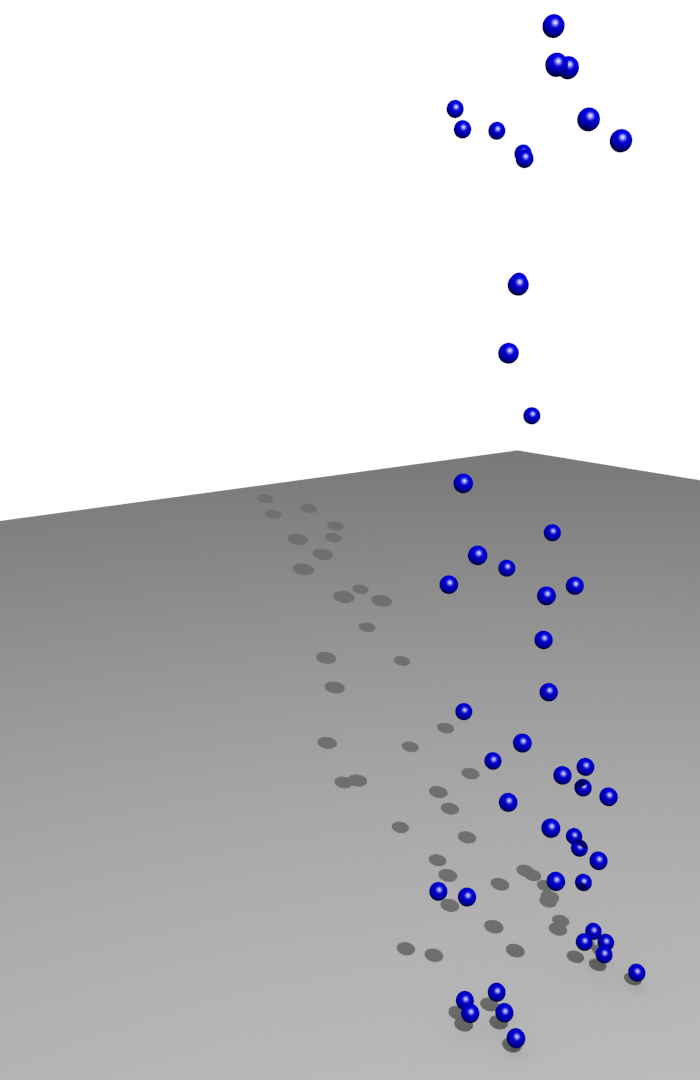}
    \includegraphics[width=0.19\textwidth]{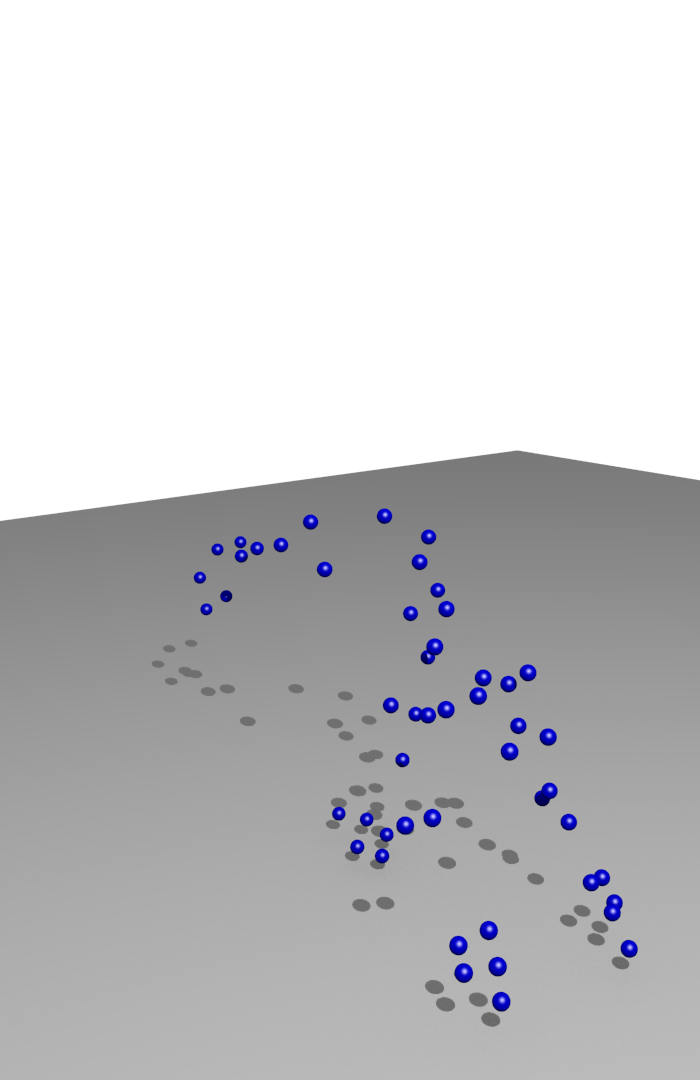}
    \includegraphics[width=0.19\textwidth]{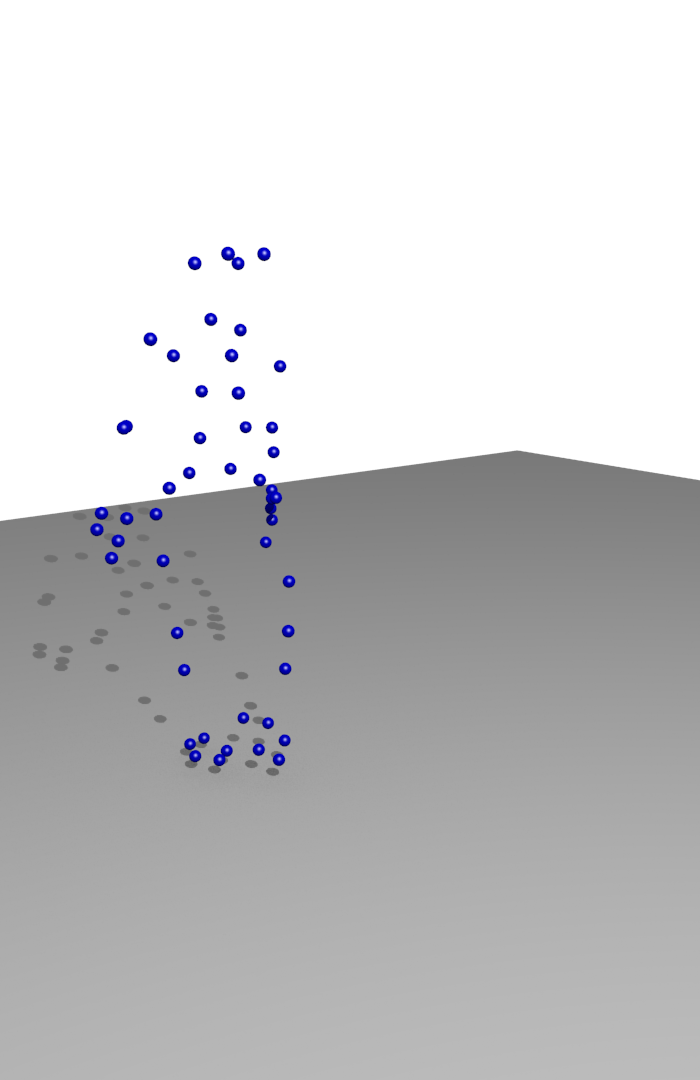}
    \includegraphics[width=0.19\textwidth]{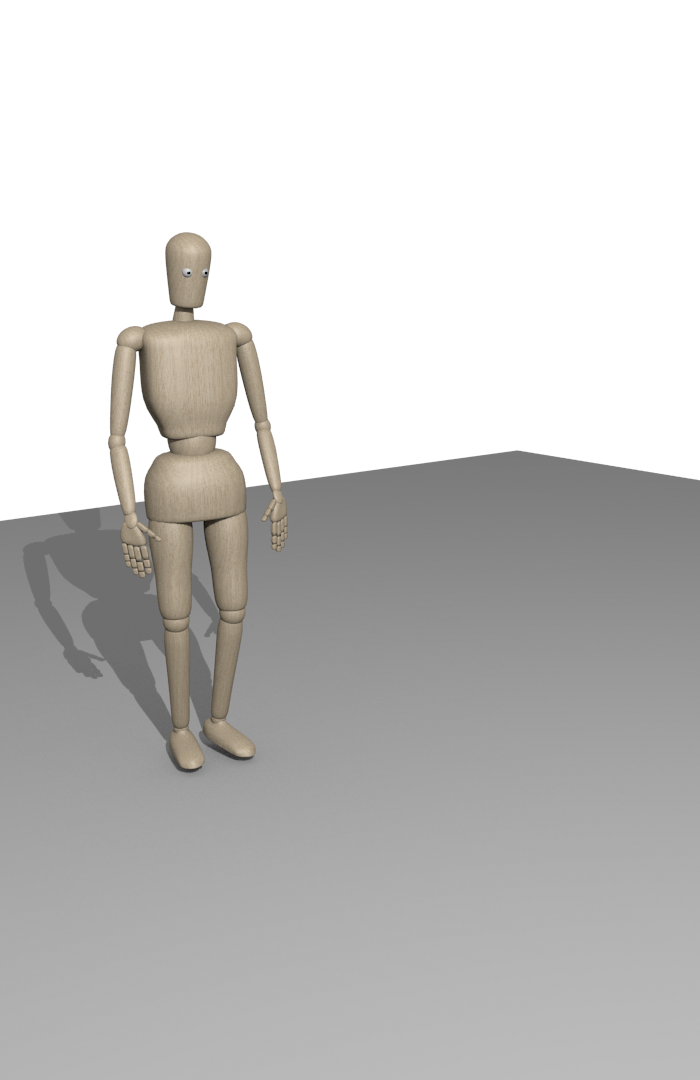}
    \includegraphics[width=0.19\textwidth]{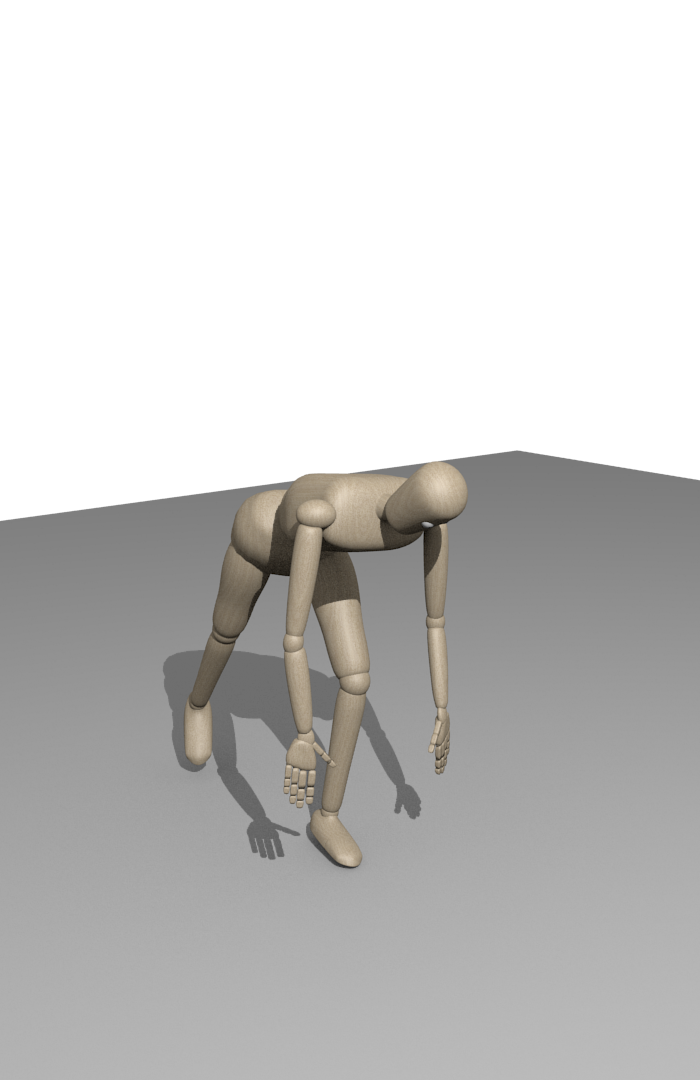}
    \includegraphics[width=0.19\textwidth]{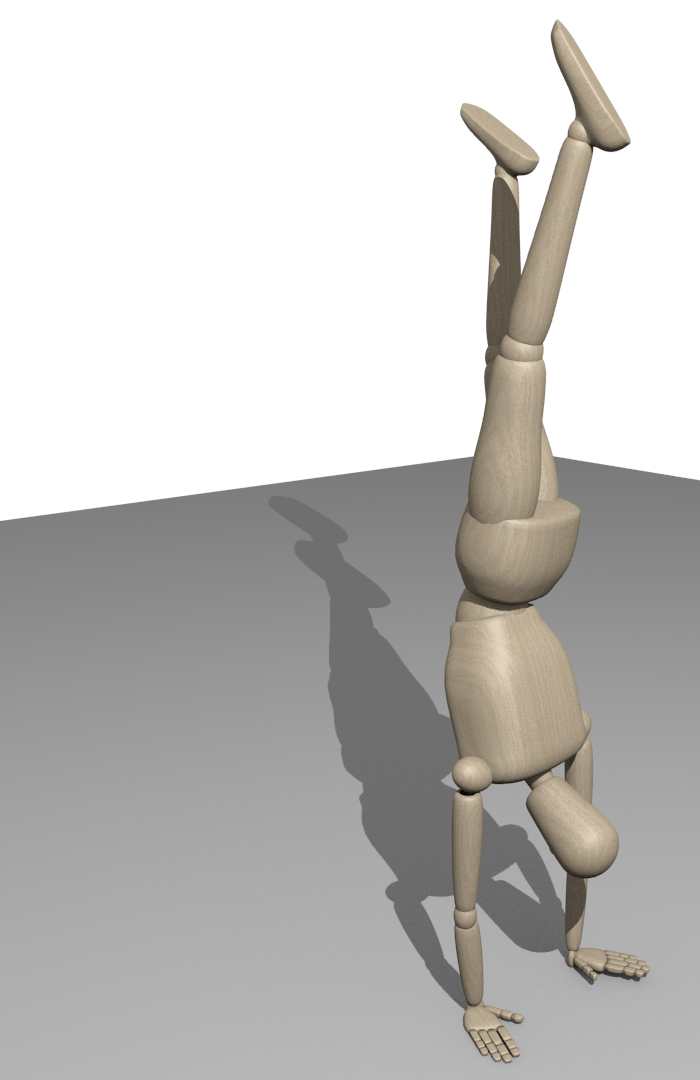}
    \includegraphics[width=0.19\textwidth]{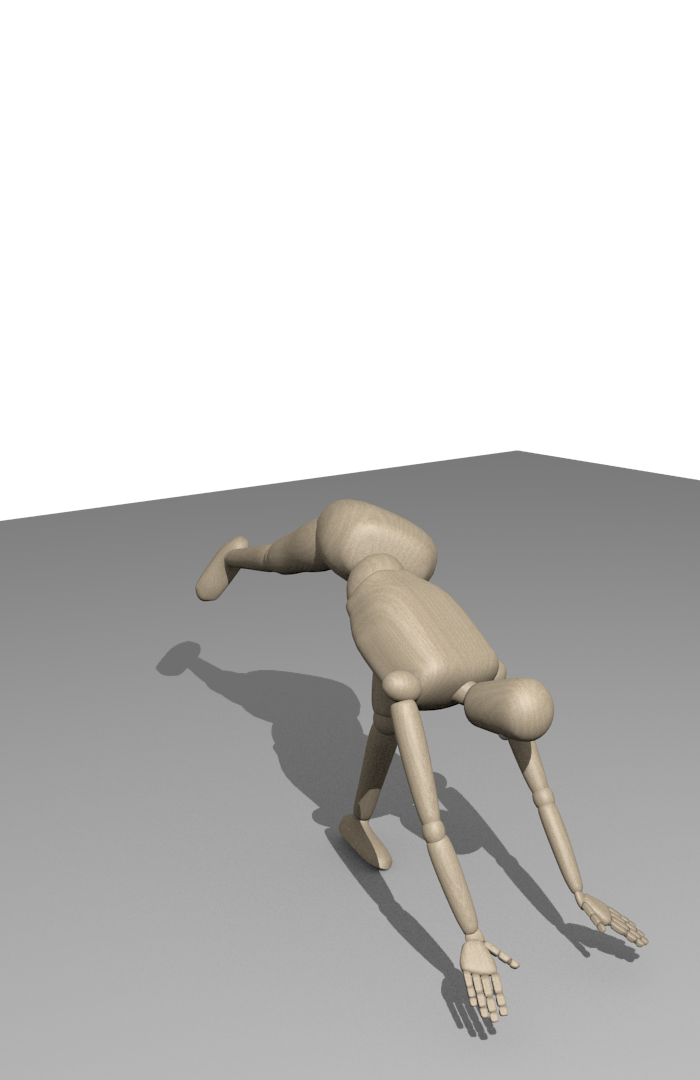}
    \includegraphics[width=0.19\textwidth]{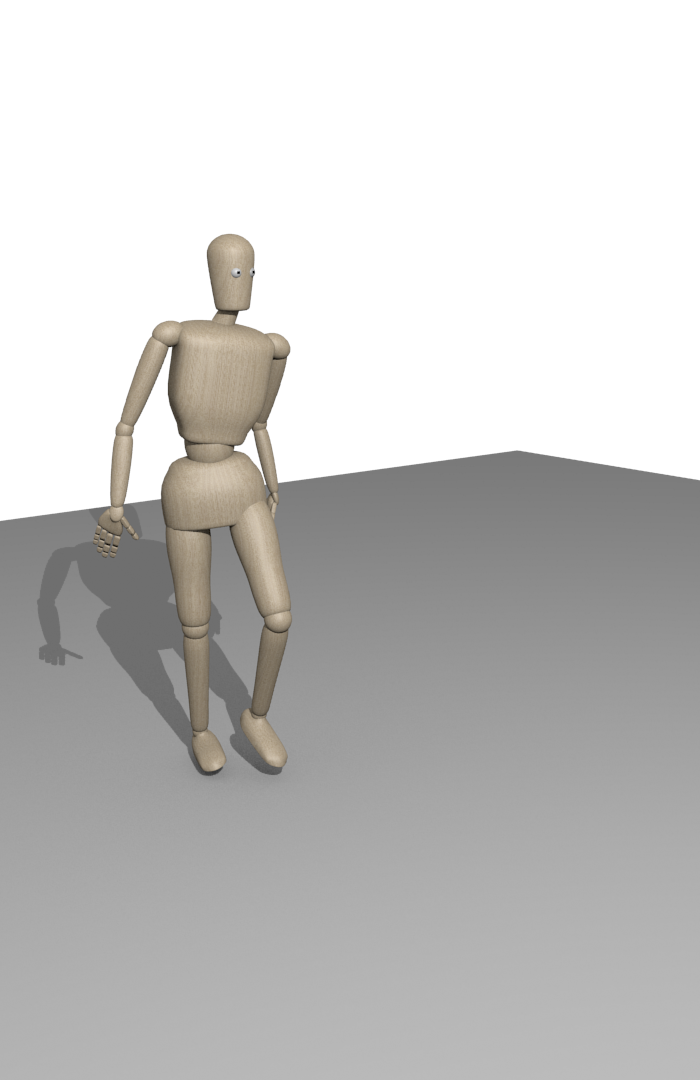}
    \caption{Selected frames of a performed handstand. In the top row, measured markers. In the bottom row, corresponding poses estimated by the proposed approach.}
    \label{fig:handstand-no-gaps}
\end{figure*}

In \Fig{fig:handstand-no-gaps-runtimes}, the runtime of the new approach is shown.
Its runtime varies over time but stays always below 8~ms.
As the Vicon systems records at 100~Hz, the proposed approach can be used to track a whole-body human motion in real-time.
Moreover, \Fig{fig:handstand-no-gaps} illustrates the recorded marker positions and the corresponding poses estimated by the new approach for selected frames.
Despite the short runtimes, the handstand is tracked very accurately.

\begin{figure*}
    \centering
    \includegraphics[width=0.19\textwidth]{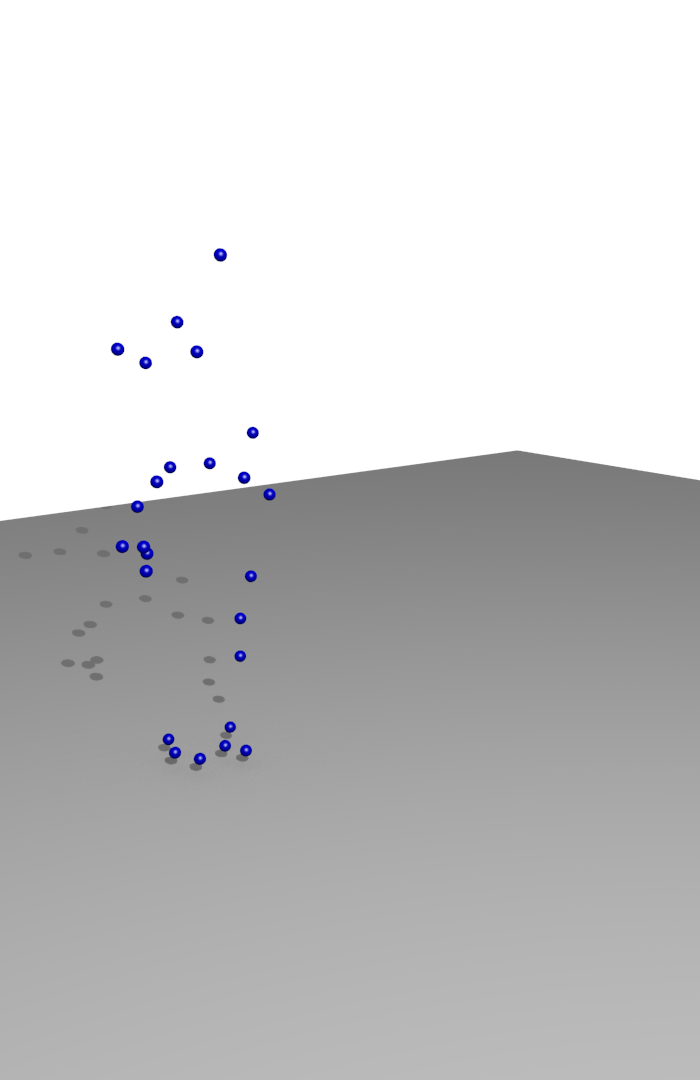}
    \includegraphics[width=0.19\textwidth]{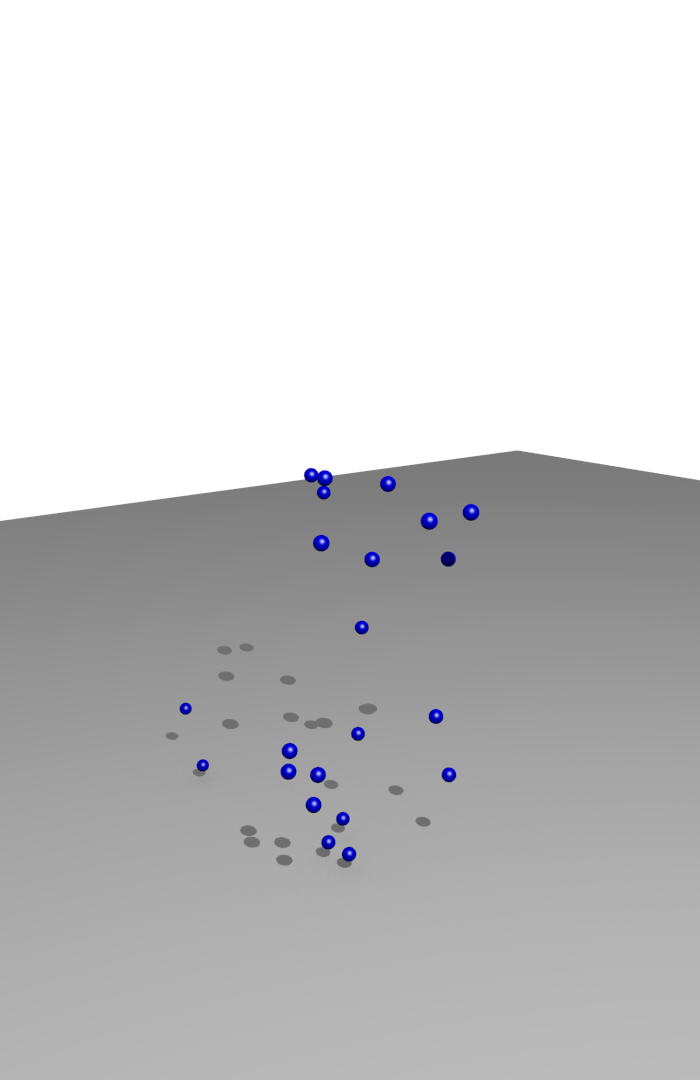}
    \includegraphics[width=0.19\textwidth]{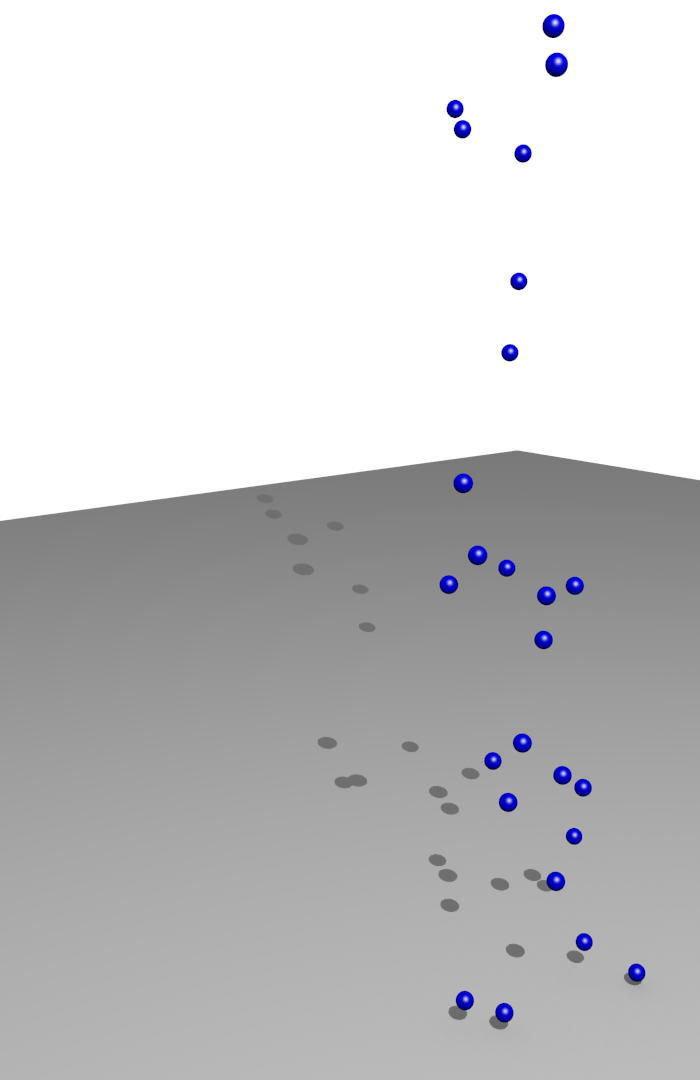}
    \includegraphics[width=0.19\textwidth]{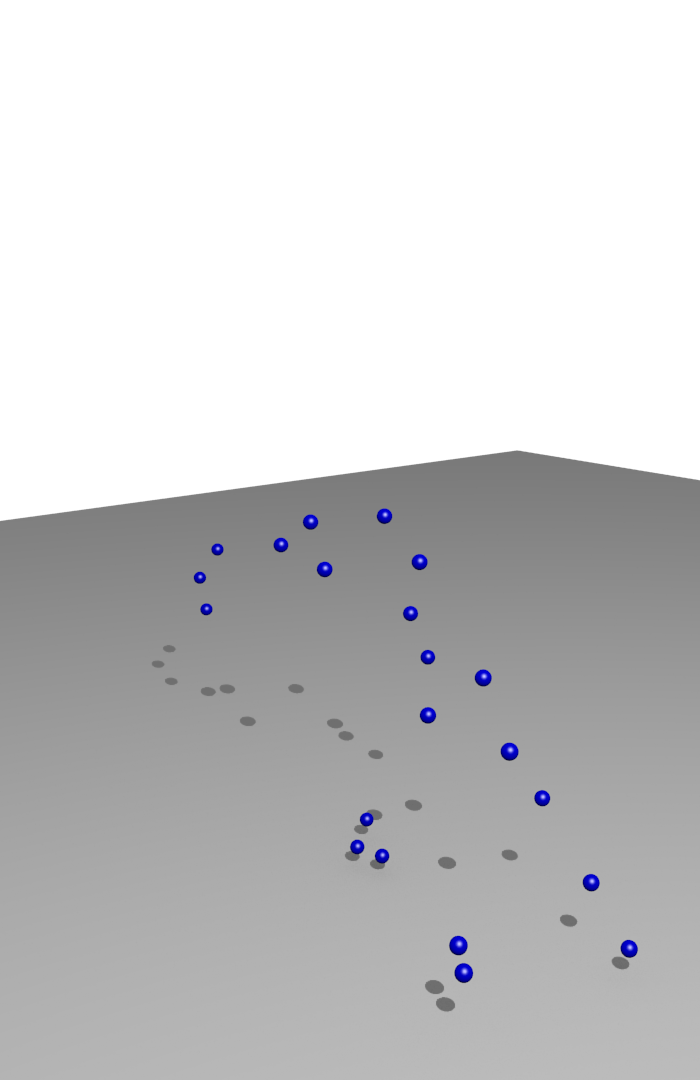}
    \includegraphics[width=0.19\textwidth]{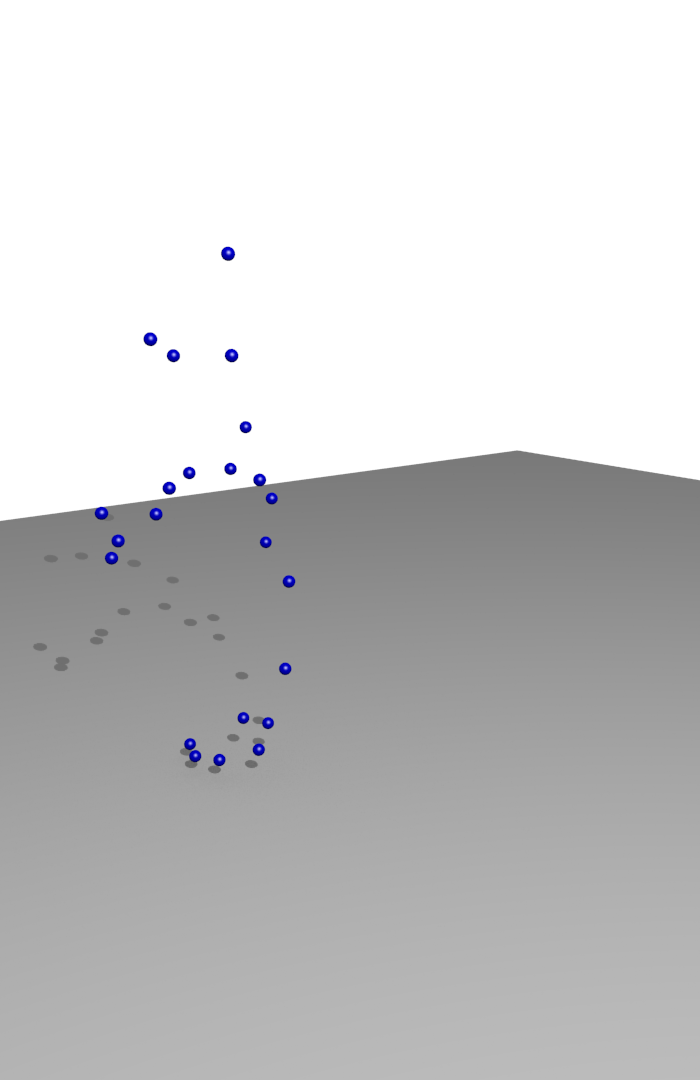}
    \includegraphics[width=0.19\textwidth]{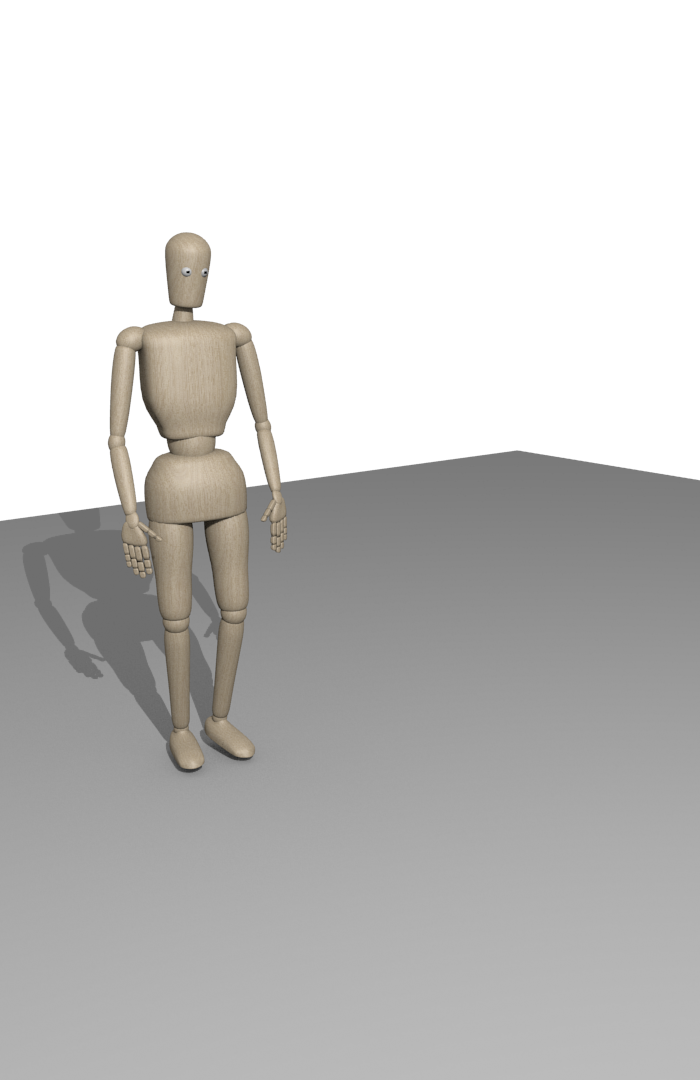}
    \includegraphics[width=0.19\textwidth]{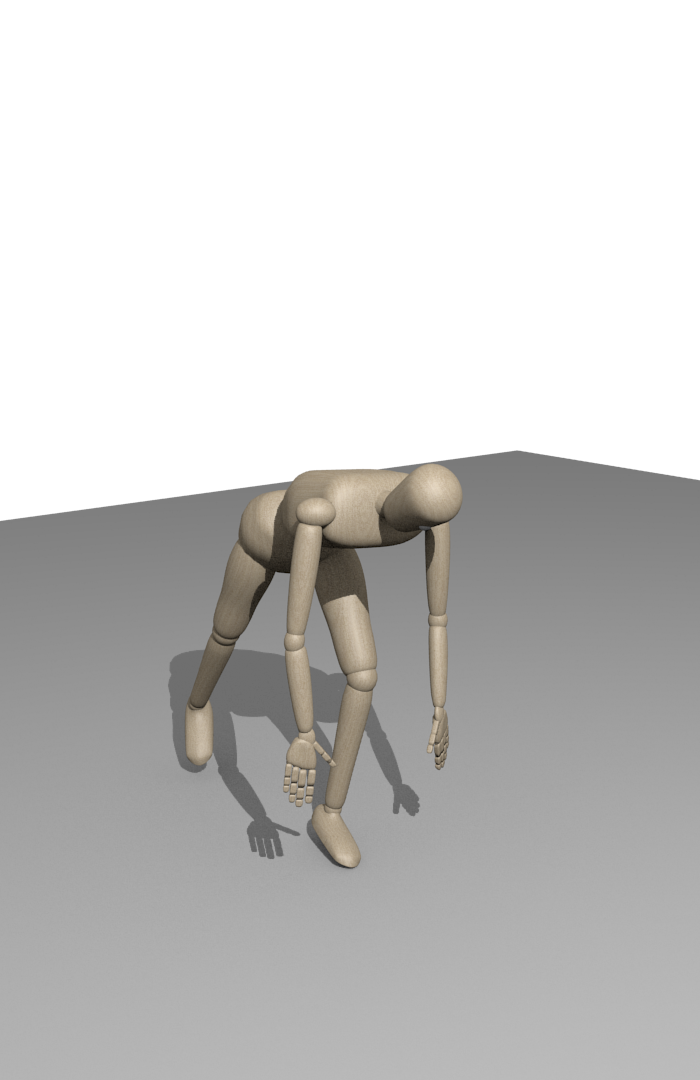}
    \includegraphics[width=0.19\textwidth]{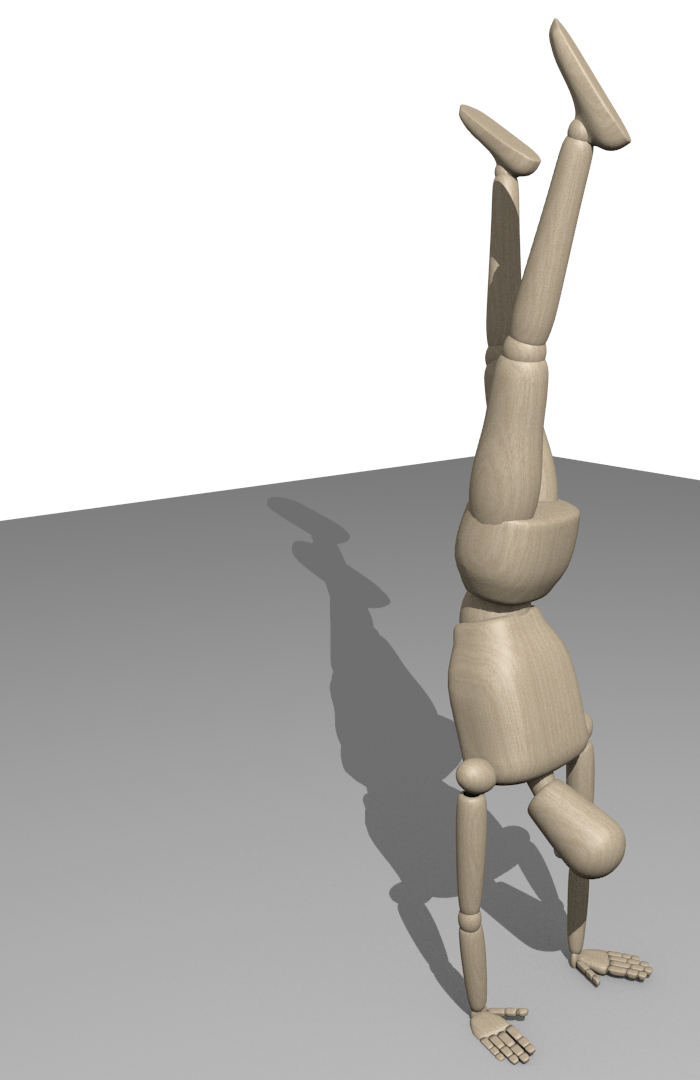}
    \includegraphics[width=0.19\textwidth]{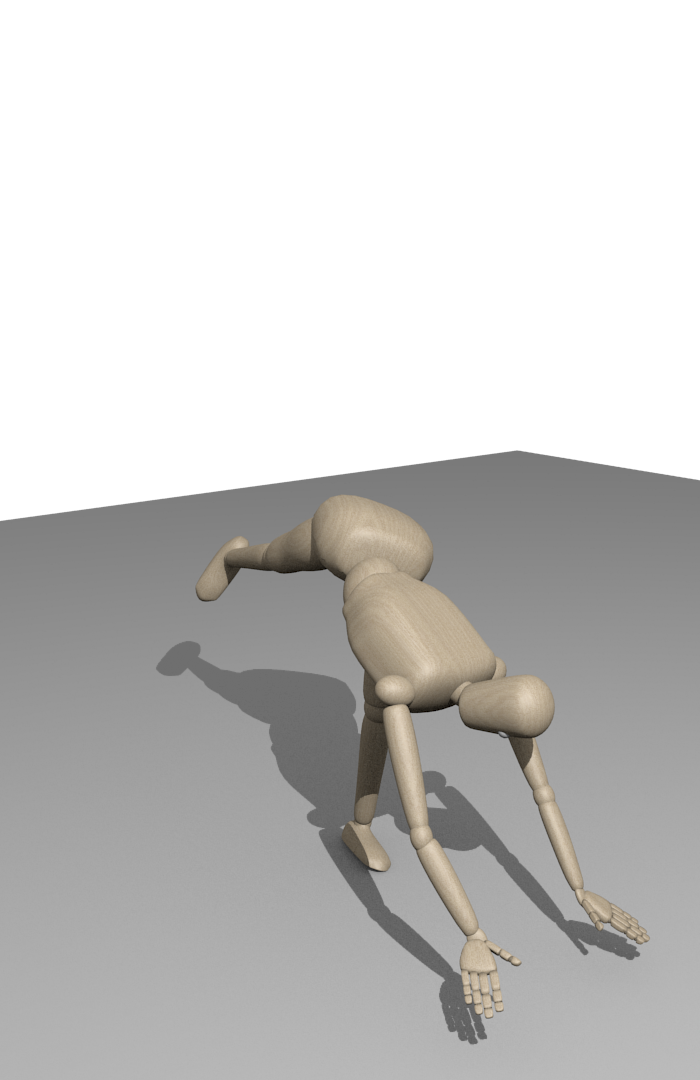}
    \includegraphics[width=0.19\textwidth]{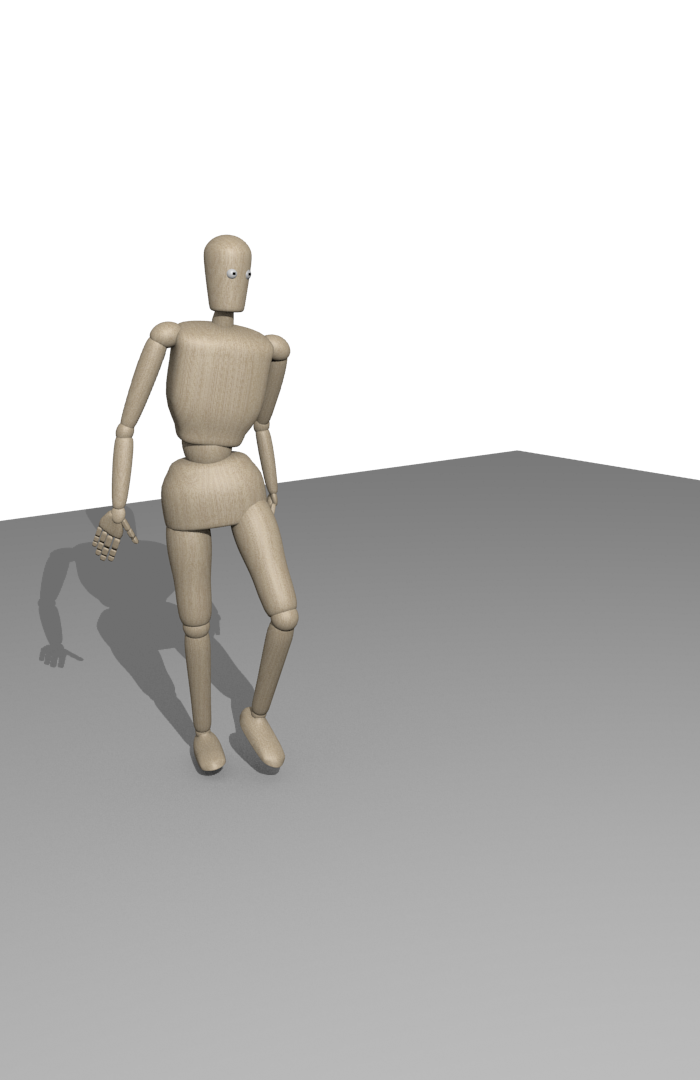}
    \caption{Same frames as in \Fig{fig:handstand-no-gaps}, but with hidden markers. In the top row, measured markers, where hidden markers are not shown. In the bottom row, corresponding poses estimated by the proposed approach.}
    \label{fig:handstand-gaps}
\end{figure*}

\begin{figure}
    \centering
    \includegraphics[width=0.55\textwidth]{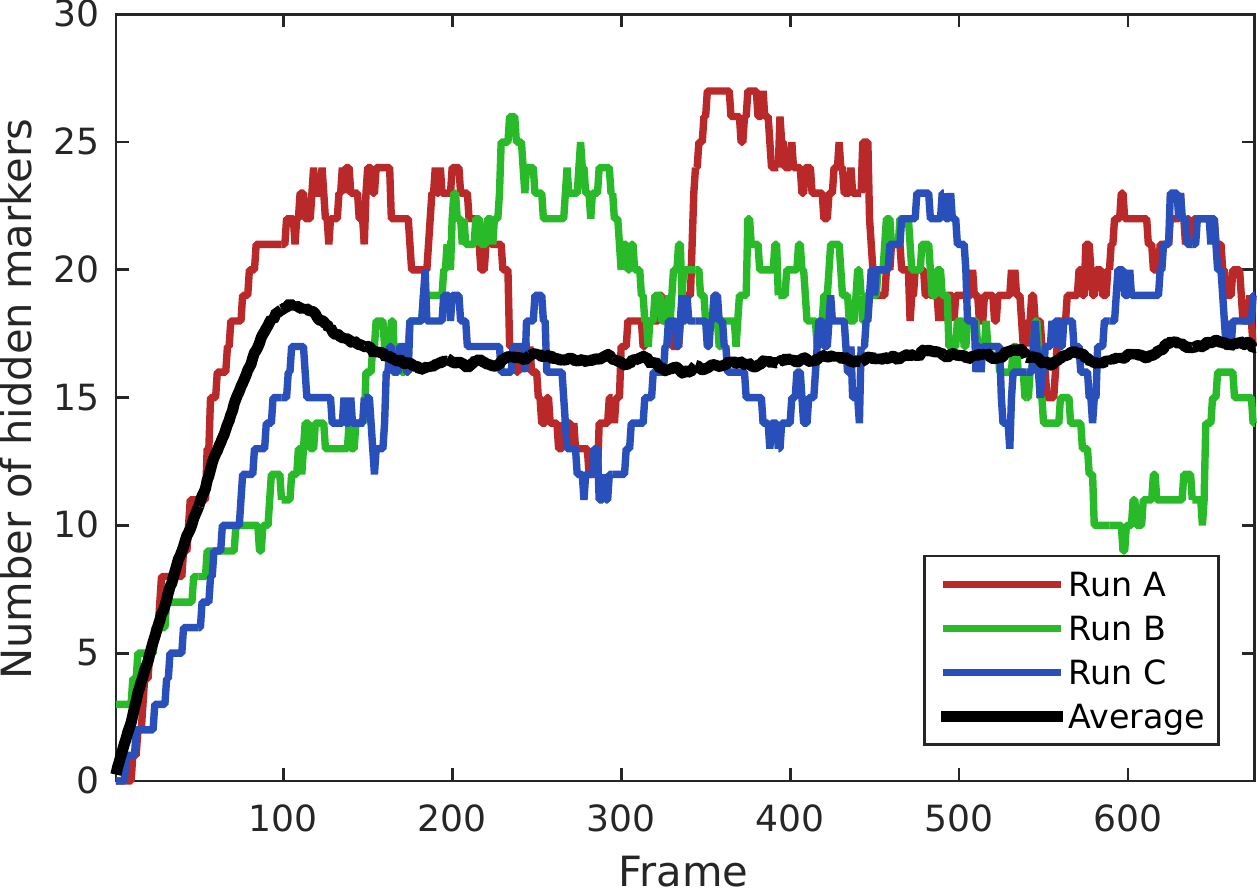}
    \caption{Number of simulated hidden markers.}
    \label{fig:handstand-unobservable-markers}
\end{figure}

In contrast to the Jacobian-based approach, the new approach can handle incomplete measurement sets as described in \Sec{sec:incomplete-meas-sets}.
To simulate markers that cannot be observed by the Vicon system for some period of time, we randomly remove markers from the previously used handstand recording.
More precisely, in every frame each marker will become invisible for the next frames with a probability of 0.5~\%.
The number of frames is determined by drawing a random number from a Poisson distribution with parameter $\lambda = 100$, i.e., on average a marker will be hidden for 100~frames (one second).
Hence, over time the total number of hidden markers will vary from frame to frame.
In order to get meaningful results, we simulate this in 100 Monte Carlo runs and try to track the human motion in each run with the randomly modified marker sets.
\Fig{fig:handstand-unobservable-markers} depicts the number of hidden markers over time for three different simulation runs as well as the average over all runs.
As can be seen, there are frames where more than 25 of the 50 markers are not available for the motion estimation.
Averaged over all runs and frames, 16 markers cannot be observed.

Although markers are assumed to be hidden for the estimation, we still know the originally recorded position.
Hence, we can compute the same marker distances as described above (also for the markers that were not available for the state estimation).
The results (blue area) are also given in \Fig{fig:handstand-marker-distances}.
The upper bound of the blue area denotes the largest averaged marker distances occurred for a run, whereas the lower bound denotes the smallest averaged marker distance.
We see that the missed markers can slightly increase the distance up to 9~cm.
Nonetheless, the results are still very good.
Looking at the motion estimated from one simulation run in \Fig{fig:handstand-gaps}, including the remaining markers available for processing, we see that the motion is very similar to the one in \Fig{fig:handstand-no-gaps}.

\section{Conclusions}
\label{sec:conclusions}

In this paper, we proposed a new way to track whole-body human motion using measurements from labeled markers attached to the human body.
Tracking a human motion is equivalent to estimating the state of a stochastic dynamic system.
Hence, we chose to rely on the Smart Sampling Kalman Filter (\sskf) to perform the human motion tracking.
This recursive state estimation approach makes it possible to systematically take the uncertainty of the marker measurements into account while being at the same time very robust to the partial collusion of markers.
However, before we could apply the filter, we had to incorporate the joint limits imposed by the human body into the estimation procedure.
This was done by transforming the constraint estimation problem into an unconstrained problem using periodic functions.
An implementation of the proposed approach was built around the kinematic reference model of the Master Motor Map and a Vicon motion capture system.
The evaluations showed that the proposed approach offers highly accurate estimates of complex whole-body human motions, even if half of the markers could not be observed.

\section*{Acknowledgment}

The research leading to these results has received funding from the European Union Seventh Framework Programme under grant agreement no 611909 (KoroiBot) and the European Union H2020 Programme under grant agreement no 643666 (I-SUPPORT).

    \bibliographystyle{IEEEtran}
    \bibliography{LiteratureH2T,LiteratureISAS}
\end{document}